\documentclass[12pt]{article}

\usepackage{float}
\usepackage{amsfonts}
\usepackage{bm}
\usepackage{amsmath}
\usepackage{amscd}
\usepackage{amssymb,lscape}

\newcommand{\bmat}{\left(\begin{array}}
\newcommand{\emat}{\end{array}\right)}
\def\gtrsim{\mathrel{\raise.3ex\hbox{$>$\kern-.75em\lower1ex\hbox{$\sim$}}
}
}

\def\ap{\alpha^{\prime}}

\def\-{\hphantom{-}}

\def\s2{\frac{1}{\sqrt2}}

\def\beq{\begin{equation}}
\def\eeq{\end{equation}}
\def\beqa{\begin{eqnarray}}
\def\eeqa{\end{eqnarray}}

\def\mg{m_{3/2}}
\def\mg2{m^2_{3/2}}

\def\Dsl{\,\raise.15ex\hbox{/}\mkern-13.5mu D} 

\def\be{\begin{equation}}
\def\ee{\end{equation}}
\def\bea{\begin{eqnarray}}
\def\eea{\end{eqnarray}}

\newcommand{\nn}{\nonumber}

\makeatletter
\@addtoreset{equation}{section}
\makeatother

\begin{document}
\pagestyle{plain}
\begin{titlepage}
\begin{center}
  \LARGE{ Double Field 
Theory description of Heterotic gauge symmetry enhancing-breaking 
\\[6mm]}
\large{\bf  G. Aldazabal${}^{a,b,c}$, E. Andr\'es$^{a,c}$
, M.Mayo${}^{a,b}$, V. Penas${}^{a,b}$
 \\}
\small{ ${}^a$ {\em G. F\'isica CAB-CNEA, }\\{\em Centro At\'omico 
Bariloche, Av. Bustillo 9500, Bariloche, 
Argentina.}\\ 
${}^b${CONICET} \\
${}^c${\em Instituto Balseiro} \\[-0.3em]
{\em  Av. Bustillo 9500,  Bariloche, 
Argentina.}\\[-0.3em]}
\end{center}

\noindent
 {\bf Abstract}:
 A Double Field Theory (DFT) description of gauge symmetry enhancing-breaking
 in the heterotic string is presented. 
 The construction, based on previous results for the bosonic string, relies on 
the extension of the tangent frame of DFT.
 The  fluxes of a Scherk-Schwarz like 
generalized toroidal  compactification are moduli dependent and become 
identified with the structure constants of the enhanced group at fixed 
``self-dual'' points in moduli space.
Slight displacements  from such points provide the breaking of the 
symmetry, gauge bosons acquiring masses proportional to fluxes.
The inclusion of fermions is also  discussed.

\vspace{.5cm}
\today


\end{titlepage}


\begin{small}
\tableofcontents
\end{small}

\newpage\section{Introduction}
\label{sec:Introduction}

 The possibility of 
understanding gauge symmetry enhancement from a Double Field Theory (DFT) 
perspective was addressed in various recent articles 
\cite{agimnr,Cagnacci:2017ulc,aamr}. The discussion was done in the context of 
the bosonic string since, even if ill defined, it is the simplest example in 
several aspects and allows to identify the  relevant ingredients. 
In the present note we follow similar steps as in \cite{aamr} in order to 
describe the gauge symmetry enhancement (and breaking) in the heterotic string 
from a DFT-like formulation. 

Gauge symmetry enhancement is a very stringy phenomena associated to the 
fact that the string is an extended object and, therefore, it can wind around 
non-contractible cycles. String states are thus characterized by a 
stringy quantum number, the so-called 
winding number, counting the number of times that the cycle is wrapped by the 
string. The exchange of winding and momentum states (accompanied by a 
transformation of  moduli fields) leads to T-duality invariance, a genuine 
stringy feature. 

 At certain moduli points (fixed points of T-duality transformations) vector 
boson 
states in some combinations of windings and momenta  become massless and  give 
rise to enhanced gauge symmetries (see for instance 
\cite{nsw,Giveon:1994fu}). Of course, the effective low energy 
theory, where massive states are neglected, can be described by an 
usual gauge field theory  Lagrangian, containing gravity, with no 
reference to 
any windings. 
An intriguing  aspect is 
that this field theory somehow  encodes information about stringy effects. 
Moreover, even if gauge symmetry breaking is achieved as usual, with 
some  scalar fields acquiring vevs, this higgsing process must encode 
information about moduli away from the fixed point.

Interestingly enough, this effective theory  
close to self-dual points originated in the bosonic string, can be 
embedded  \cite{aamr} into a DFT-like formulation. 
In DFT (we will be more precise below) the internal configuration space 
includes, besides the usual space coordinates dual to KK momenta, new 
coordinates dual to winding states and therefore, coordinates are 
doubled. 
This DFT rewriting allows to highlight the stringy aspects of these 
gauge theories. Actually, in a  
generalized 
Scherk-Schwarz \cite{ss,effective} compactification of this DFT  the 
 fluxes, computed from an internal vielbein depending on doubled 
coordinates, appear to depend on moduli and become the structure constants of 
the enhanced group at fixed points.
We show below that this rewriting also works for the bosonic sector of a 
toroidally compactified  heterotic string.
Moreover, we show that by invoking supersymmetry,  a corresponding 
fermionic 
sector can also be  introduced.

In Section \ref{sec:DFT rewriting} we present a brief discussion of 
symmetry enhancement and show  the DFT rewriting of heterotic string theory 
effective action close or at the enhancing points. 
It  is also shown how breaking of 
gauge symmetry is  encoded into the moduli dependence of fluxes. 
A simple illustration for the  case of  circle 
compactification is provided. Ideas presented in \cite{aamr} are 
recurrently  used 
throughout the article.

  The introduction 
of fermions is discussed in Section \ref{sec:fermions}.
In particular we show that if the gaugings in shift matrices of gauged 
supergravities, 
associated to fermionic mass terms,  are replaced by Scherk-Schwarz (moduli 
dependent) fluxes, the masses of  fermions are in correspondence with their 
bosonic partners, as expected from 
supersymmetry.

Several details are presented in  the Appendices. In Appendix
\ref{sec:Heterotic DFT}
a quick introduction to DFT and generalized Scherk-Schwarz like 
compactification is provided with emphasis in the heterotic case where the 
ingredients   needed in our construction are highlighted. For a 
more 
complete introduction to DFT we provide some original 
references in \cite{orDFT} and refer the reader to some reviews 
\cite{reviewamn,gm,reviews} (where a more extensive list of references can be 
found).
In Appendix \ref{sec:Heterotic string basics} a brief account of 
heterotic 
string features needed for our discussion is presented.

Concluding remarks and a brief  outlook are presented in Section
\ref{sec:Summary and outlook}.

\section{Heterotic Gauge symmetry enhancement and DFT rewriting}
\label{sec:DFT rewriting}
Toroidal compactification of the $SO(32)$ (or $E_8\times E_8$) heterotic string 
to $d$ space-time 
dimensions leads to a generic gauge group 
\begin{equation}
 G_L\times U(1)^{10-d}_R
\end{equation}
where the left group $G_L$ is generically a product of non-abelian and abelian 
gauge groups. The rank of  $G_L$ is  $r_L=16+10-d=26-d$ originated from 
the $16$ Cartan 
generators of the ten dimensional gauge group plus the $r=10-d$ 
vector bosons coming from  left combinations  of the KK  reductions of 
the metric and the antisymmetric tensor.
Different gauge groups do appear when moving along moduli space. At 
generic points in moduli space $G_L=  U(1)^{26-d}_L$ while a point 
of maximum enhancement leads to $G_L= SO(52-2d)$ for the $SO(32)$ string case.

We present some basic details in Appendix \ref{sec:Heterotic string 
basics}. 
Let $n=n_c+r_L=\dim{G}_L$ be the dimension of $G_L$ at some moduli point with 
$n_c$ denoting the number of charged generators.
The effective low energy theory will thus be a $G_L\times U(1)^{10-d}_R$ gauge 
theory coupled to gravity and the Kalb-Ramond  antisymmetric tensor field in 
$d$ dimensions. 
There are also $(n_c+26-d)(10-d)$ scalars.
Thus, the counting of degrees of freedom leads to:  $d^ 
2$ corresponding 
to graviton plus B field, $n_c+36-2d$ vectors from $G_L\times U(1)^{10-d}_R$  
and $(n_c+26-d)(10-d)$ scalars.
Recall that the number of scalar fields corresponds to 
$(26-d)(10-d)$ moduli 
plus $n_c(10-d)$ extra scalars that should become massive at generic points 
where the broken gauge group is  $U(1)^{26-d}_L \times U(1)_R^{10-d}$.

It is interesting to notice that  the total
number of degrees of freedom 
coincides with 
\begin{equation}
 dim \frac{O(d+n,d+r)}{O(d+n)\times O(d+r)}=d^2 + d(n_c+36-2d)+(n_c+26-d)(10-d).
\label{dftembedding}
\end{equation}
Indeed, this coset-like writing provides a clue of how to express the effective 
theory 
in a DFT-like form as discussed in Appendix \ref{sec:Heterotic DFT}.

Following similar steps as presented in 
\cite{agimnr,aamr} for the bosonic string case,  we propose an expression for 
such 
an action and then discuss its specific features. Namely,
\bea \nn
S_{eff}&=& \frac{1}{2\kappa_d^2}\int d^dx\sqrt{g}e^{-2\varphi}
\left[
     {\cal R}+4\partial^\mu\varphi\partial_\mu\varphi-\frac1{12}H_{\mu\nu\rho}
     H^{\mu\nu\rho} \right. \\\label{actionDFTeff}
     &&\left.-\frac 18{\cal H}^{{}}_{AB}{ F}^{A\mu\nu}
{F}^B_{\mu\nu}
+\frac 18 (D_\mu {\cal H})_{AB}
(D^\mu {\cal H})^{AB}-V \right].
\eea
Here
\bea \label{scalarpotential}
V=-\frac 1{12}f_{AB}{}^Kf_{LC}{}^D\left( {\cal 
H}^{AL}{\cal 
H}^{BC}{\cal H}_{KD} \right. 
- 3\, {\cal H}^{AL} \eta^{BC} \eta_{KD} + 2 \, \eta^{AL} 
\eta^{BC} \eta_{KD}
) 
- \Lambda
\eea
is a scalar potential where the last two terms are just constants. 
The scalars parametrize the  coset $\frac{O(n,r)}{O(n)\times O(r)}$ of 
dimension $(n_c+26-d)(10-d)$.
The  indices can be  conveniently split in a  L-R basis (named a  $\cal 
C$ 
 base) as 
 $A=(a,\hat{I})$ where  $a=1,\dots r_L,r_L+1,\dots r_L+n_c=n=\dim{G}_L$   
index 
runs over the  left group $G_{L}$. In addition the  $\hat{I}=1,\dots r$  
index corresponds 
to the Right $U(1)^r$  group.
The index contractions are performed with $ \eta^{AB}$, the $O(r_L+n_c, 
r)$ 
invariant metric 
  \be \label{etam}
\eta^{AB}=\begin{pmatrix} 1_{r_L+n_c} & 0 \\ 0 & -1_r \end{pmatrix}.
\ee

${\cal H}_{AB}$ is the (so-called)  internal generalized metric encoding 
information about scalar fields. 
${\cal R}$ is the $d$-dimensional Ricci scalar and 
$F_{\mu\nu}^A$ and
$H_{\mu\nu\rho}$ 
\bea \label{FHDFT}
F^B & = & d A^B - \frac{1}{2\sqrt{2}}f_{CD}{}^B A^C \wedge A^D \nn
\\
H   & = & d B + F^C \wedge A_C- \frac{1}{3!\sqrt{2}} f_{ABC}A^A\wedge A^B \wedge A^C,
  \eea
are the gauge field and $B$ field strengths.

The covariant derivative of the scalars is
  \be 
  (D_\mu {\cal H})_{AB}=(\partial_\mu {\cal H})_{AB}+
\frac{1}{\sqrt{2}}f^K{}_{LA} A^L_\mu
     {\cal H}^{{}}_{KB} + \frac{1}{\sqrt{2}}f^K{}_{LB} A^L_\mu{\cal 
H}^{{}}_{AK}.
\,
\label{dftcovder}
  \ee
 Finally, the   $ f_{ABC}= \eta_{AK}f^K{}_{BC}$ are 
completely antisymmetric constants. 
Interestingly enough this action  can be  interpreted as a generalized 
Scherk-Schwarz reduction of a DFT-like action, as we briefly sketch in
Appendix \ref{sec:Heterotic DFT}, the constants  $ f^K{}_{BC}$ 
being the generalized fluxes of 
the compactification\footnote{Recall that other kinds of fluxes like $f_A$ 
could be present \cite{effective, reviewamn,Dibitetto:2012rk} as shown in 
Appendix \ref{sec:Heterotic DFT}. Here we set  them 
to 
zero  since they are not relevant for our discussion.}. There are 
$\frac{(r+n)(r-1+n)(r-2+n)}{3!}$ 
such fluxes which must satisfy the quadratic constraints
\begin{equation}
f_{[AB}{}^K
f_{K]C}{}^R=0\, .
\label{quadratic}
\end{equation}
If indices are allowed to transform then the  
action is globally invariant under $O(n_c+26-d,10-d)$
and it can be identified with  the bosonic (electric) sector  of a 
half-maximal gauged 
supergravity 
action \cite{Schon:2006kz, 
Samtleben:2008pe,Trigiante:2016mnt,Bergshoeff:1985ms}.

In spite of the fact that this huge number 
of  gaugings was explored in several situations, its physical interpretation 
deserves  further investigation. 
For instance, if we restricted to $a=1,\dots, r=10-d$, and in 
$r=6$  
dimensions, the  above counting of fluxes  would correspond to the 
$220$ gaugings of 
electric sector of $O(6,6)$ gauged supergravity. These gaugings  have been 
identified (see for instance \cite{Samtleben:2008pe,fluxes, reviewamn} ) as  
geometric and 
non geometric fluxes in (orientifold) string 
compactifications. Here  we will restrict to a very specific choice of a subset 
of all possible fluxes, relevant to our discussion.

In order to make contact with the heterotic string effective action we first 
expand the generalized metric in terms of scalar fluctuations encoded in the 
 scalar  matrix $M_{a,\hat I}$ with $dim G_L\times  r=(n_c+26-d)(10-d) $ 
independent degrees of freedom. Namely we 
write 
\begin{equation}
 {\cal H}_{\cal C}^{AB}= \delta^{AB}+ {\cal H}^{(1)AB}+ \frac12{\cal 
H}^{(2)AB}+\dots
\label{scalarfluct}
\end{equation}
  such that matrix elements  vanish unless
\begin{eqnarray}
 {\cal H}^{(1)}_{a\hat I}&=& M_{a,\hat I},\qquad{\cal H}^{(1)}_{\hat I 
a}=M^T_{a,\hat I}\\\nn
{\cal 
H}^{(2)}_{ab}&=& (M M^T)_{ab},\qquad{\cal 
H}^{(2)}_{\hat I \hat J}= (M^T M)_{\hat I \hat J}\, .
\label{scalarorders}
\end{eqnarray}

 Moreover, we make a specific choice for flux values (therefore breaking the 
global symmetry), by identifying them with the gauge group structure constants.
Namely, 

\be
f_{ABC}=\begin{cases} f_{abc}\qquad G_L\,\, \rm {structure\,\, constants} \\
\nonumber 
0\qquad \rm {otherwise} 
\end{cases} 
\label{Gl},
\ee
where  $f_{abc}$ is the subset of all possible fluxes (with Left indices) 
reproducing the  structure constants of the $G_L$ group algebra.
When couplings are adequately adjusted the above action reduces to the  
$G_L\times U(1)^{10-d}_R$ gauge theory action 
\bea \label{effectiveSD}\nn
S&=& \frac{1}{2\kappa_d^2}\int d^dx\sqrt{g}e^{-2\varphi}
\left(
{\cal
R}+4\partial^\mu\varphi\partial_\mu\varphi-\frac1{12}H_{\mu\nu\rho}H^{\mu\nu\rho
}\right) \\\nn &-&\frac 18 \left(\delta_{ab}{F}^{a\mu\nu}{F}^b_{\mu\nu} + 
\delta_{\hat I\hat J}
{\bar F}^{\hat I\mu\nu} { \bar F}^{\hat I}_{\mu\nu} -\frac 12 g_d\sqrt{\ap} 
M_{a\hat I} 
F^a_{\mu\nu} \bar F^{\hat I\mu\nu }\right) \nn\\
 &-&
 D_\mu M_{a\hat I}D_\nu M^{a\hat I} g^{\mu\nu}
 +
 {\cal
 O}(M^4),
 \label{su2action}
\eea
reproducing the bosonic sector of  
heterotic string  low energy   theory at a fixed point. 
Here $a$ labels the Left gauge group  (generically non- Abelian) 
generators 
with vector bosons $ A_{L\mu}^a$  and $\hat I=1,\dots r$ the Abelian group 
$U(1)_{\hat I}$  associated to vector bosons $ A_{R\mu}^{\hat I}$. 
The scalar fields live in  the $(\bf {dim G_L})_{\hat q =0}$ adjoint 
representation of $G_L$ and carry zero vector charge  ${\bf {\hat q }}=(\hat 
q_1,\dots,\hat q_r)=0$ with respect to $U(1)_{R}^r$ right group.
Thus, the covariant derivative in \eqref{dftcovder} becomes
\begin{equation}
 D_\mu M_{a\hat I}=\partial_\mu M_{a\hat I}+
g_d f^k{}_{la} A_{L\mu}^l
     M^{{}}_{k\hat I}\, ,
\end{equation}
where $g_d=\kappa_{d}\sqrt{\frac{2}{\ap}}  $.
Notice that no scalar potential is generated for this choice of structure 
constants.

In the next section we show, in the context of DFT, how gauge symmetry breaking 
can 
be achieved by allowing structure constants to depend on moduli, 
as expected from string theory.

\subsection{ Gauge symmetry breaking from  DFT rewriting}
\label{sec:GSBDFT rewriting}

In string theory the structure constants can be read out from 3-point vertex 
vector 
boson operators. For the Cartan generators, which  we label with the index 
$\check{I}_{L}=(i=1,\dots r;I=1,\dots 16$), the associated Left vector 
bosons vertex operators are of the form
$V(\check{I}_{L})\propto \partial_z y^{\check{I}} \tilde\psi^{\mu}e^{iK.X}$, 
whereas for 
charged operators we have $V({l_L})\propto e^{i l_L.y(z)}e^{iK.X}$ where 
$l_L^{\check{I}}$ are the Left internal momenta defined in 
\eqref{leftrightmomenta}, 
and $X^{\mu}(z)$ and $ K^{\mu}$ are the space-time coordinate and momentum, 
respectively. 
Recall that the internal  momenta depend on specific values of 
KK momenta 
$p_m$ and winding numbers $\tilde p^m$ as well as  on the 
$\Lambda_{16}$
weights \footnote{For the sake of 
clarity we concentrate in the $SO(32)$ string but same conclusions are valid 
for 
the $E_8\times E_8$ heterotic case with lattice  $\Lambda_{8}\times 
\Lambda_{8}$ .} $P^I$. We encode these values into a ``generalized momentum 
vector''  
\begin{equation}
 \check {\mathbb P}=(\mathbb P; P^I),\qquad {\rm with }\qquad \mathbb 
P=(p_m,{\tilde p}^m).
\label{generalizedmomentum}
\end{equation}

Let us encode denote by $\Phi=(g,B,A)$ a generic 
moduli point. Since  
momenta depend on moduli fields we 
actually have  $l_L=l_L^{\check 
{\mathbb P}}(\Phi)$ and similarly, $l_R=k_R^{\check 
{\mathbb P}}(\Phi)$.

At  specific points $\Phi_0$ in moduli space and for certain values of $\check 
{\mathbb P}$ such that 
\begin{equation}
 k_R ^{\check{\mathbb P}}(\Phi_0)=0,\qquad (l_L^{\check {\mathbb 
P}}(\Phi_0))^2=2
\label{masslessvector}
\end{equation}
 gauge symmetry enhancement occurs 
\eqref{masslesscharged}. At these points 
\begin{equation}
 l_L^{({\check{\mathbb P})}}(\Phi_0)\equiv \alpha^{({\check{\mathbb P})}}
 \label{momentumroot}
\end{equation}
become the roots  $\alpha^{({\check{\mathbb P})}} $  of the  $G_L$ gauge group. 
Notice that there is an 
associated root to each of the $n_c$ possible  values  of  ${\check{\mathbb 
P}}$, 
satisfying the massless vector condition \eqref{masslessvector}. 
 
Three point amplitudes involving Left vector boson vertices can be expressed 
as 
\begin{eqnarray}
\langle 
V_L({l_L^{(\check{\mathbb P}_1)}})
V_L({l_L^{(\check{\mathbb P}_2)}})
V_L({l_L^{(\check{\mathbb P}_3)}})
\rangle
 &\propto & f_{ \alpha^{(\check{\mathbb P}_1)}\alpha^{(\check{\mathbb 
P}_2)}\alpha^{(\check{\mathbb P}_3)}(\Phi)}E(\epsilon_1,K_1; 
\epsilon_2,K_2;\epsilon_3,K_3)\, ,
\end{eqnarray}
where $E(\epsilon_i,K_i;i=1,2,3)$ is a Lorentz invariant  antisymmetric 
function of polarization vectors $\epsilon_i^{\mu}(K_i)$ and space-time momenta 
$K_i^{\mu}$. The constants $f_{ \alpha^{(\check{\mathbb 
P}_1)}\alpha^{(\check{\mathbb P}_2)}\alpha^{(\check{\mathbb P}_3)}}(\Phi)$ 
are antisymmetric and  vanish unless internal momentum is conserved, namely  
$\check{\mathbb P}_3=-\check{\mathbb P}_1-\check{\mathbb P}_2$.
At a  self-dual 
point $\Phi=\Phi_0$ this indicates  that structure constants $f_{ 
\alpha_1\alpha_2\alpha_3}$ vanish unless  $\alpha_1+\alpha_2$ is a root . In 
this case, we can normalize by setting $f_{ 
\alpha^{(\check{\mathbb 
P}_1))}\alpha^{(\check{\mathbb P}_2)}\alpha^{(\check{\mathbb 
P}_3)}}(\Phi)=1$.
Momentum conservation also implies that, at the self-dual point, amplitudes 
mixing Left and Right indices vanish. However, away from the fixed point, the 
vertices develop a dependence on $l_{R}$,  
$V_L(l=({l_L,l_{R})})\propto e^{i l_L^{\check 
{\mathbb P}}(\Phi).y(z)+l_R^{\check 
{\mathbb P}}(\Phi).\bar y(\bar z) }e^{iK.X}$ and therefore mixing now occurs. 
In 
fact,  it is found that
 the only non vanishing amplitudes are  
\begin{eqnarray}\nn
\langle V_L({l^{(\check{\mathbb P})}})
V_L({l^{(-\check{\mathbb P})}})
V_{L}(\check{I})\rangle
&\propto & l_{L}^{(\check{\mathbb 
P)}}(\Phi)_{ \check{I}};\qquad \langle V_L({l^{(\check{\mathbb P})}})
V_L({l^{(-\check{\mathbb P})}})
V_{R}(\hat{I})\rangle
\propto  l_{R}^{(\check{\mathbb 
P)}}(\Phi)_{ \hat{I}}.
\label{structureconstvertices}
\end{eqnarray}
Following \cite{aamr},  we propose to identify the amplitude coefficients with 
some algebra 
structure constants, even (slightly) away from the fixed point $\Phi_0$. 
Namely we set
\begin{eqnarray}
f_{ 
\alpha^{(\check{\mathbb P})}\alpha^{(-\check{\mathbb 
P})}{ \check{I}_L}}(\Phi)&=& l_{L}^{(\mathbb 
P)}(\Phi)_{ \check{I}}
,\qquad  f_{ \alpha^{(\mathbb 
P)}\alpha^{(\check{\mathbb 
-P)}}\hat I}(\Phi)= l_{R}^{(\check{\mathbb 
P)}}(\Phi)_{\hat I}
\end{eqnarray}
with the other constants being obtained as permutations, and we propose the 
algebra
\bea\nn
 \big[ E_{\alpha},E_{-\alpha } \big] &=& l_{L}^{(\alpha)\check  I} 
H_{\check  I}+l_{R}^{(\alpha)\hat I}\hat H_{\hat I} 
\qquad   \big[ H_{I},E_{\alpha } \big] = l_{L}^{(\alpha)I} E_{\alpha}       
   \\
 \big[ E_{\alpha_1},E_{\alpha_2 } \big] &=&f_{ 
\alpha_1\alpha_2\alpha_3} E_{\alpha_3 }\qquad 
     \big[ \hat H_{ I},E_{\alpha } \big] = l_{R}^{(\alpha)I} 
E_{\alpha}\, .
\label{generalalgebra}
\eea
We have used $\alpha=\alpha^{(\mathbb 
P)}$  to alleviate the notation and, as we found above, $f_{ 
\alpha_1\alpha_2\alpha_3}=1 $ if $\alpha_3=\alpha_1+\alpha_2$ is a root  and 
vanishes otherwise. All other commutators vanish.
It is easy to show that \eqref{generalalgebra} satisfies  Jacobi identities 
and, therefore, defines a  Lie algebra.

Recall that, at the self-dual point where  $k_{R}^{\alpha}(\Phi_0)=0 $ and $f_{ 
\alpha -\alpha \hat I}=  l_{L}^{(\alpha)\hat I}=\alpha 
^{\hat I}$,  the algebra reduces to  to the 
gauge algebra of $G_L$ group in the Cartan-Weyl basis.
For instance  
$[E_{\alpha},E_{-\alpha}]=\alpha^ {{\hat I}}H_{\hat 
I}$ for charged generators $E_{\alpha}$ and Cartan generators $H_{\hat I}$, as 
expected.
Interestingly enough, by performing a linear combination of generators it can 
be shown that there is still an underlying  $G_L$ algebra. To visualize 
the linear combination let us define a double Cartan operator 
$\mathbb{H}_{A}=\big(H_{\check I} , \hat H_{ \hat{I}}  \big)$ and the  double 
(moduli dependent) momentum $\mathbb{L}^{(\alpha)}_{A}=\big( 
l_{L}^{(\alpha)\check{I}}, 
l_{R}^{(\alpha)I} \big)$. The algebra \eqref{generalalgebra} can now be 
written as
\begin{equation}
\begin{aligned}
\big[E_{\alpha},E_{-\alpha} \big]&= \mathbb{L}^{(\alpha)}_{A}\mathbb{H}_{A} \\
\big[\mathbb{H}_{A},E_{\alpha} \big]&= \mathbb{L}^{(\alpha)}_{A} E_{\alpha} \\
\big[E_{\alpha_{1}},E_{\alpha_{2}} \big]&= f_{\alpha_{1}\alpha_{2}\alpha_{3}} 
E_{\alpha_{3}}
\end{aligned}\, .
\label{generaloddalgebraleftright}
\end{equation}

It is worth observing that an  $O(r_{L},r_{R})$ transformation can be 
performed over the double Cartan 
generator, namely the one given by the inverse of 
\begin{equation}
\begin{pmatrix}
\delta_{IJ} && 0 && A_{Jn}\delta_{nm}\\
-A^{J}_{m}\delta_{JI} && \delta_{nm} && G_{nm}-B_{nm} 
-\frac{R}{2}A^{I}_{n}A^{I}_{m}\\
-A^{J}_{m}\delta_{JI} && \delta_{nm} && -G_{nm}-B_{nm} 
-\frac{R}{2}A^{I}_{n}A^{I}_{m}
\end{pmatrix}\, ,
\label{isomorphismalgebra}
\end{equation}
such that  $\mathbb{L}^{(\alpha)}$ is mapped  to $\check {\mathbb P}$ (see 
\eqref{generalizedmomentum}) and 
$\mathbb{H}$ to 
new double Cartan's $\mathcal{H}$ leading to  
\begin{equation}
\begin{aligned}
\big[E_{\alpha},E_{-\alpha} \big]&= 
\check{\mathbb{P}}^{(\alpha)}_{A}\mathcal{H}_{A} \\
\big[\mathcal{H}_{A},E_{\alpha} \big]&= \check{\mathbb{P}}^{(\alpha)}_{A} 
E_{\alpha} \\
\big[E_{\alpha_{1}},E_{\alpha_{2}} \big]&= f_{\alpha_{1}\alpha_{2}\alpha_{3}} 
E_{\alpha_{3}}.
\end{aligned}
\label{generaloddalgebra}
\end{equation}
This final algebra has the same form independently of moduli values.  
Furthermore since the  
algebra \eqref{generalalgebra} 
 and \eqref{generaloddalgebra} are isomorphic, due to 
\eqref{isomorphismalgebra}, we conclude  that the algebra at the 
self-dual point is the same at all other (neighborhood) points.

In generalized Scherk-Schwarz like compactifications of DFT,  the generalized 
fluxes $f_{ABC}$ are defined from the generalized algebra  satisfied by the 
internal frame \eqref{structureconstants}. Let us assume for the moment that a 
specific  choice 
of frame exists such that these fluxes are the structure constants found in 
\eqref{generalalgebra}. Once these fluxes are identified we must replace 
them into the action \eqref{actionDFTeff}. The output is that the resulting 
action 
is the gauge broken symmetry action where vector bosons and scalars acquire 
masses proportional to structure constants mixing left and right indices, 
namely 
$
 f_{ \alpha^{(\mathbb 
P)}\alpha^{(-\mathbb P)}\hat I}(\Phi)
$. 
 \subsubsection{Goldstone bosons} 
We  start by inspecting the couplings 
between vectors and scalars arising from kinetic terms in \eqref{actionDFTeff}. 
By 
keeping  the 
first term in the internal metric expansion \eqref{scalarfluct}, 
 ${\cal H}_{\cal C}^{AB}= \delta^{AB}+ {\cal 
H}^{(1)AB}+\dots$ we find that
\begin{eqnarray}
D_\mu {\cal H}_{AB}
D^\mu {\cal H}^{AB}&\approx& 4\partial_\mu 
M^{AB}f^K{}_{LA}\delta_{KB}A^{\mu L}=4 \partial_\mu 
M^{a\hat I}f_{aL \hat I}A^{\mu L}\\&=&
4 \partial_\mu 
M^{a \hat I}f_{a b \hat I}A^{\mu b}\, .
\end{eqnarray}
Here we have used the exapnsion   into Left and Right 
indices $A=(a,\hat{I})$,  we have used the metric  \eqref{etam}, the 
antisymmetry of  $f_{ABC}$ and the 
fact that the only non vanishing fluxes are of the form $f_{abc}, f_{ab \hat 
I}$. 

The conclusion is that,  for a given vector boson $A_{\mu}^{ b}$, 
there 
is a combination of $ \hat I=1,\dots,r=10-d$, would-be Goldstone bosons  
scalar fields $f_{a b \hat I} M^{a \hat I}\equiv f_{\alpha -\alpha  \hat I} 
M^{\alpha \hat I}= l_{R}^{(\mathbb 
P)}(\Phi)_{ \hat{I}} M^{\alpha \hat I}$ ( whenever $f_{a b 
\hat I}\ne 0$).  We have recast the expression in a Cartan-Weyl basis
by recalling 
that the only non vanishing fluxes (away from the point of enhancement) 
containing a Cartan index are of the form 
$f_{\alpha -\alpha \hat I}$.

Interestingly enough, this  combination  arises as 
a conformal anomaly contribution in the OPE of energy momentum tensor  
with  scalars  whenever these scalars become massive, away from the fixed point 
(see \cite{agimnr} for a bosonic 
string example).  
This indicates that the combination  $ l_{R}^{(\mathbb 
P)}(\Phi)_{ \hat{I}} M^{\alpha \hat I}(K)$ of internal R-momentum and scalar 
polarizations, must be set to zero.\footnote{Alternatively, such a combination 
of scalar vertex operators must be included into a new massive, anomaly free,  
vector field.}  

Let us see, as an example, how vector bosons and scalar masses arise.

\subsubsection{Vector masses} 
In order to read the vector boson masses we must just look at quadratic 
terms in the scalar kinetic term. Thus, following similar steps as above 
but now keeping just the constant term in the internal metric expansion 
\eqref{scalarfluct}  ${\cal 
H}_{AB}= \delta_{AB}+\dots$ we find
\begin{eqnarray}\nn
\frac 18 (D_\mu {\cal H})_{AB}
(D^\mu {\cal H})^{AB}&\approx& 
\frac 18 (f_{RLB}\delta_{KA}+f_{RLA}\delta_{KB})\eta^{RK}
(f^{PSA}\delta^{P'B}+f^{PSB}\delta^{P'A})\eta_{PP'}A_{\mu}^LA^{\mu}_S\\\nn
&=& 2 \frac 18  (
f_{RLB}f^{PSA}\delta^{P'B}+f_{RLB}f^{PSB}\delta^{P'A} 
)\delta_{KA}\eta_{PP'}\eta^{RK}A_{\mu}^LA^{\mu}_S\\\nn
&=& -\frac 12  f_{ \hat I aL }f^{ \hat I a S }A_{\mu}^LA^{\mu}_S
= -(f_{\alpha 
-\alpha \hat I}(\Phi))^2 |A^{\alpha}|^2 \, ,
\label{goldstonebosons}
\end{eqnarray}
where, again, a Cartan-Weyl rewriting was used in the last term.
Namely, away from the fixed point, the vector bosons acquire a mass 
$m_{A^\alpha}$ given by
\begin{equation}
 m_{A^\alpha}^2= \sum_{\hat{I}}( f_{\alpha -\alpha\hat{I}}(\Phi) 
)^{2}=l_{L}^2(\Phi) \, .
\end{equation}

\subsubsection{Scalar masses}

From a  DFT point of view, the scalar  masses arise from quadratic terms in 
scalar fluctuations in the scalar potential. Thus, 
by inserting the expansion \eqref{scalarfluct} into the scalar potential 
\eqref{scalarpotential} we find:
\begin{eqnarray}
-\frac{1}{12}f_{ABC}f_{DEF}\mathcal{H}^{(1)AD}\mathcal{H}^{(1)BE}\delta^{CF}
-\frac{1}{12} f_ {ABC}f_{DEF} \mathcal{H}^{(2)AD}\big(3\delta^{BE}\delta^{CF} - 
3\eta^{BE}\eta^{CF} \big)\, .
 \end{eqnarray}
We notice that, due to the relative minus sign between left and right indices 
in $\eta^{AB}$ (see\eqref{etam})  the second term vanishes unless indices 
organize as $\delta^{be}\delta^{\hat I \hat J}$ leading to
\begin{equation}
\frac{1}{2}\sum_{\alpha,\hat{I}}f_{\alpha -\alpha\hat{I}}f_{\alpha-\alpha 
\hat{I}}\mathcal{H}^{(2)\alpha \alpha}=\frac{1}{4}\sum_{\alpha,\hat{I},\hat{J}} 
( f_{\alpha -\alpha\hat{I}} )^{2} | M^{\alpha \hat{J}}|^{2}=
\frac{1}{4}\sum_{\alpha,\hat{J}} m_{\alpha}^2 |M^{\alpha \hat{J}}|^{2}\, ,
\end{equation}  
where 
\begin{equation}
 m_{\alpha}^2 =\sum_{\hat{I}}( f_{\alpha -\alpha\hat{I}} 
(\Phi))^{2}= m_{A^\alpha}^2
\end{equation}
is the mass (square) of the scalar field  $M^{\alpha \hat{J}}$, coinciding 
with the vector boson mass.

 On the other hand, the first term  contribution in 
\eqref{goldstonebosons} leads to \begin{equation}
\frac{1}{2}\sum_{\alpha} \Big(\sum_{\hat{I}} f_{\alpha -\alpha \hat{I}} 
m^{\alpha \hat{I}} \Big)^{2}.
\end{equation}
However,  this contribution is irrelevant since  $\Big(\sum_{\hat{I}} f_{\alpha 
-\alpha \hat{I}} m^{\alpha \hat{I}} \Big)$ is the Goldstone boson combination.

Let us stress that the 
obtained masses
coincide with the masses computed from string mass formula 
\eqref{LRstringmasses}.

\subsection{Examples}
Here we discuss a simple  illustration of the above construction in the 
simplest case of  compactification on a circle of radio $R$.
 In this case \eqref{leftrightmomenta} reads
\bea
K_L^I&=& P^I+ R A^I \tilde p\\\nn
k_{L}&=&\sqrt{\frac{\ap}{2}}\big[\frac{p}{R}+\frac{\tilde p}{\tilde R}-P.A 
-\frac{R}{2} A. A\tilde p\big]\\\nn
k_{R}&=&\sqrt{\frac{\ap}{2}}\big[\frac{p}{R}-\frac{\tilde p}{\tilde 
R}-P.A-\frac{R}{2}A.A\tilde p \big]\, .
\label{leftrightmomentaexam}
\eea
A massless state requires $k_{R}=0$ ($\bar{N}_{F}=1$) and then 
$k_{L}=\sqrt{2\ap}\frac{\tilde p}{\tilde R}$.

\subsubsection{$SU(2)\times SO(32)\times U(1)$ }
 A possible set of massless vectors is provided by  
choosing $p=\tilde{p}=\pm 1$, by  setting the radio to its self-dual value 
$R=\sqrt{\alpha'}=\tilde R$ and  $A^I=0$. Together with the 
massless vector state associated to KK compactification (with $p=\tilde p=0=P^ 
I$)  mode these massless vectors lead to an $SU(2)$ left group. 

In addition, an $SO(32)$ group  associated to 
the  weights $P=( \underline {\pm,\pm,\dots 0})$ appears (underlining meaning 
permutation over the 16 entries) and the corresponding 16 Cartan 
oscillators. 
Therefore, at this moduli point, the enhanced gauge group is  
$SO(32)_{L}\times SU(2)_{L}\times U(1)_{R}$. 
In the notation of 
\eqref{generaloddalgebraleftright} this  full set of massless states 
corresponds 
to 
$\check {\mathbb P}_{SU(2)\times 
SO(32)\times U(1)}=(\pm 1,\pm 1;  0,\dots ;0),(0,0;\underline 
{\pm,\pm,\dots 0})$.

We can break the enhanced gauge group $SO(32)_{L}\times SU(2)_{L}\times 
U(1)_{R}$ to $SO(32)_{L}\times U(1)_{L}\times U(1)_{R}$ or to 
$U(1)^{17}_{L}\times U(1)_{R}$ depending on the direction of the moduli space 
on 
which we move. 

For instance, by sliding away from the self-dual radio, 
charged $SU(2)$ vectors become massive with mass square $m_{-}^2$ 
with $m_{-}=\sqrt{\frac{2}{\ap}}a_-=\frac{1}{R}-\frac{1}{\tilde{R}} $, where 
$a_{\pm}$ are defined in \eqref{amp}.
The algebra 
\eqref{generaloddalgebraleftright} becomes
\begin{eqnarray}
\big[E_{+},E_{-} \big]&=&2( a_{+} H_{3} + a_{-}H_{\bar{3}})\hspace{3em}
\big[E_{P},E_{-P} \big]=P_{I}H_{I} \nonumber\\
\big[H_{3},E_{\pm}\big]&=&\pm a_{+} E_{\pm}\hspace{8em}
\big[H_{I},E_{P} \big]=P_{I}E_{P} \nonumber\\
\big[H_{\bar{3}},E_{\pm}\big]&=&\pm a_{-} E_{\pm}
\hspace{7em}
\big[E_{P_{1}},E_{P_{2}} \big]= f_{P_{1}P_{2}P_{3}} E_{P_{3}}\, .
\end{eqnarray}
The subindices $\pm$ denote the two roots of $SU(2)$, the subindex $3$ denotes
the corresponding Cartan whereas $f_{P_{1}P_{2}P_{3}}$ are the structure 
constants of $SO(32)$ where $P_{I}$ are the  roots and $H_{I}$ the Cartan 
generators. At the self-dual radio we have $a_-=0$, $a_+=1$  and the $ 
SU(2)$ gauge algebra is recovered.

By turning on Wilson lines $A^{I}$ the group is broken to  
$U(1)^{17}_{L}\times U(1)_{R}$. 
The algebra becomes
{\footnotesize{
\begin{eqnarray}
\big[E_{+},E_{-} \big]&=&\big(2-\frac{1}{2}A^{2}\big)H_{3} + 
\big(\frac{1}{2}A^{2}\big) H_{\bar{3}}+ A^{I}H_{I}
\hspace{2em}
\big[E_{P},E_{-P} \big]=P_{I}H_{I} -(P\cdot A)H_{3} -(P\cdot A)H_{\bar{3}} 
\nonumber\\
\big[H_{3},E_{\pm}\big]&=&\pm \big(2-\frac{1}{2}A^{2}\big) E_{\pm}
\hspace{10,5em}
\big[H_{I},E_{P} \big]=P_{I}E_{P} \nonumber\\
\big[H_{\bar{3}},E_{\pm}\big]&=&\pm \big(\frac{1}{2}A^{2}\big) E_{\pm}
\hspace{11,5em}
\big[E_{P_{1}},E_{P_{2}} \big]= f_{P_{1}P_{2}P_{3}} E_{P_{3}}\nonumber\\
\big[H_{I},E_{\pm}\big]&=&\pm A^{I} E_{\pm}
\hspace{13,5em}
\big[H_{\bar{3}},E_{P} \big] =-(P\cdot A )E_{P} \, .
\end{eqnarray}}}
As discussed, the vector boson masses are identified with the structure 
constants mixing Left and Right indices. Therefore we find that  $SU(2)$ 
charged 
vectors $A_{\mu}^{\pm}$  acquire a mass 
$m_{SU(2)}=|f_{\bar3\pm}{}^{\pm}|=\frac{1}{2}A^{2}$
 whereas $SO(32)$ charged vectors   masses are  $m_{SO(32)}=|f_{\bar 
3 P -P}|= |P\cdot A|$. 
As discussed in the general case, the above commutators satisfy Jacobi 
Identities and define an $SU(2)\times SO(32)$ algebra now involving massive 
states. 
Let us recall that from DFT perspective the algebra is obtained through 
generalized Lie derivatives of the twists $E_A(\mathbb Y)$. The explicit 
twist for the $SU(2)$ sector is given in \eqref{su2frame}.

\subsubsection{$SO(34)$ }
Other enhanced groups can be obtained at different points in moduli space. 
points in moduli space can lead to different enhancements. For instance, 
by  choosing \cite{nsw} $\tilde R= \sqrt{2\ap}$ and $RA=(-1,\dots 0)$ we
notice that for $\tilde p=0$ massless states are obtained if   
$ P=(\underline {\pm,\pm,\dots 0})$, namely an   $SO(32)$ root,  if  KK 
momenta  ${p}=-P^1$ is selected. 
Moreover, the $SO(32)$ weights $P=(\pm,\underline {\pm,\dots 0}), (0,\dots 
0), (2,\dots 0)$ combined with $\tilde p=\pm1$ lead to  
$l_L=(\pm; \underline {\pm,\dots 0})$ states that combined with the $SO(32)$ 
roots 
lead to massless states with charged operators associated to $l_L=(\underline 
{\pm;\pm \dots 0})$ corresponding to the well known $SO(34)$ enhanced group 
\cite{nsw}.

Recall thet our description holds at the neighborhood of the $SU(2)\times 
SO(32)\times U(1)$ point (defined by a specific choice of generalized momentum 
$\check{\mathbb P}_{SU(2)\times 
SO(32)\times U(1)}$ and moduli fields) or  $SO(34)$ point with  different 
generalized momenta $\check{\mathbb P}_{SO(34)}$ and moduli fields but it is 
not possible to continuously interpolete between both points.
\section{Including fermions}
\label{sec:fermions}

The action \eqref{actionDFTeff}, for 
$d=4$,   is nothing but the $N=4$ bosonic (electric) sector of a generic gauged 
supergravity theory (see for instance 
\cite{Bergshoeff:2002nv,Schon:2006kz,Samtleben:2008pe,Trigiante:2016mnt}).
We then see that, Scherk-Schwarz reduction of DFT provides a way of deriving 
this 
gauged supergravity sector.

Inclusion of the magnetic sector requires considering 
EFT or an extension of the initial global group.
The inclusion of fermions from a DFT point of view was considered in several 
works \cite{dftfermions} and, in particular,
a Scherk- Schwarz 
like 
reduction was proposed in \cite{Berman:2013cli} in the context of the 
superstring.

The aim of the present section is to show that the mechanism of gauge 
symmetry enhancing- breaking through moduli dependent  fluxes, found for the 
bosonic sector, is 
reproduced in the fermionic sector.

By invoking supersymmetry we conclude that  the fermionic 
sector is just the fermionic sector of 
gauged supergravities discussed in the literature.
 We first concentrate in the $N=4$ case in 
four dimensions and discuss its generalization later on. Therefore, we 
must deal with  the global symmetry group $O(6+n,6)$. In 
particular we concentrate  in the fermionic mass terms. For instance, 
quadratic 
terms containing the gravitini $\psi{}_{\mu i}$ and gaugini 
$\lambda^{a}_{j}$ read \cite{Bergshoeff:2002nv, Schon:2006kz} 
\begin{eqnarray}
  e^{-1} {\cal L}_{\text{f.mass}} \, &=&
                           \, \frac13 \, g \,A_{1}^{ij} \, \bar 
\psi{}_{\mu i} \, \Gamma^{\mu \nu } \, \psi_{\mu j}
 + i g \, {A_{2\,a i}}^{j}\bar 
\psi{}_{\mu i} \,\Gamma^{\mu} \, \lambda^{a}_{j}+ 
A_{3ab}{}^{ij}\bar \lambda^{a}_{j}\lambda^{a}_{j}     + 
\text{h.c.} \; ,
\end{eqnarray}
where the matrices $A_{1}^{ij}, {A_{2\,a i}}^{j}, A_{3ab}{}^{ij}$
are known as shift matrices. Here indices $i$ span the spinorial representation 
of $SO(6)$ or, equivalently,  the 
4-dimensional representation of $SU(4)$, the universal 
cover of $SO(6)$.
 $SO(6)$ vectors $v_{\hat m}$ can be recast  in terms 
of the antisymmetric combinations of spinorial representations, or, 
equivalently 
in terms of antisymmetric $SU(4)$ six dimensional representation $v^{ij}$ 
through  $v_{\hat m} (\gamma^{\hat m}){}^{ij} = v^{ij}$
where 
\begin{equation}\label{pseudorealityconst}
v^{ij}=v^{[ij]}~~~ \mbox{and}~~~ 
v_{ij}=(v^{ij})^{*}=\frac{1}{2}\epsilon_{ijkl}v^{kl}.  	
\end{equation}
The shift matrices are known to depend on scalars through the coset 
representatives ${\cal U}_{A}{}^{\bar A}(x)$ defining the scalar matrix 
\eqref{scalarmatrix}. For  internal indices such matrix reads 
\beq 
{\cal H}_{AB}(x) ={\delta }_{\bar A \bar B}\, {\cal U}_{A}{}^{\bar A}  {\cal 
U}_{B}{}^{\bar B} \ .
\label{internalscalarmatrix}\eeq
with
\begin{equation}
  {\cal U}_{A}{}^{\bar A}(x)\equiv ( {\cal U}_{A}{}^{a}; {\cal U}_{A}{}^{\hat 
I})
=(  {\cal U}_{A} {}^{a};  {\cal U}_{A}{}^{ij})\, ,
\label{cosetrep}
\end{equation}
and where the $SO(6)$ vector index $\hat I$ was expressed in terms of the 
spinor 
indices $ij$ in the last term. 

The shift matrices then read (see for 
instance\cite{Bergshoeff:2002nv, Schon:2006kz}) 
\begin{eqnarray}\nn
  \mbox{gravitini-gravitini:}~~       A_{1}^{ij}&\propto& 
( {\cal U}_{A}{}^{kl})^{*}
{\cal U}_{B}{}^{ik}{\cal U}_{C}{}^{jl}f{}^{ABC}\\\nn
\mbox{gravitini}-\mbox{gaugini}:~~
A_{2ai}{}^{j}&\propto&{\cal U}_{A}{}^{a}({\cal U}_{B}{
}^{ik})^{*}{\cal U}_{C}{}^{jk}f{}^{ABC}\\
\mbox{gaugini}-\mbox{gaugini}:~~
A_{3ab}{}^{ij}&\propto &{\cal U}_{A}{}^{a}{\cal U}_{B}{}^{b}
{{\cal U}}_{C}{}^{ij}f{}^{ABC}\, ,
\label{shiftmatrices}
\end{eqnarray}
where we have used $f{}^{ABC}$ to denote the electric sector 
gaugings $f_{+}{}^{ABC}$, the $+$ subindex indicating the electric 
sector\cite{Schon:2006kz}. In order to read vector 
masses we need to keep the constant term in the 
expansion of $ {\cal U}_{A}{}^{\bar A}$ in scalar fluctuations (see \eqref 
{internalscalarmatrix}) reproducing 
the metric  expansion   ${\cal 
H}_{AB}= \delta_{AB}+{\cal O}(M)$. Therefore   $( 
{\cal U}_{A}{}^{ b};{\cal U}_{A}{}^{\hat I})=(\delta_{A} {}^{ b}; 
\delta_{A}{}^{\hat 
I})$ with
\begin{equation}\label{vielbeindeltassu4}
{\cal U}_{A}{}^{ij}=\delta_{A \hat m}(\gamma^{\hat m})^{ij}.
\end{equation}

By replacing this expansion into shift matrices expressions we find
\begin{eqnarray}\nn
A_{1}^{ij}&\propto& 
\delta_{A,\hat a} {(\gamma^{\hat a})}^{* kl}
\delta_{B,\hat I} {(\gamma^{\hat I})}^{ik}
\delta_{C,\hat c} {(\gamma^{\hat c})}^{jl}f{}^{ABC}=
 {(\gamma^{\hat a})}^{* kl}
 {(\gamma^{\hat I})}^{ik}
 {(\gamma^{\hat c})}^{jl}f{}^{\hat a\hat I\hat c}
\\\nn
A_{2ai}{}^{j}&\propto&
 {(\gamma^{\hat I})}^{* ik}
 (\gamma^{\hat c})^{jk}f{}^{a\hat I\hat c}\\
 A_{3ab}{}^{ij}& \propto &
 {(\gamma^{\hat c})}^{ij}f{}^{ab \hat c}.
\end{eqnarray}

By identifying the gaugings $f{}^{ABC}$ with the fluxes defined above and 
by using that  fluxes involving more than one right index 
vanish  
($f_{}{}^{\hat a\hat I\hat c}=f_{}{}^{a\hat I\hat c}=0$) we find that, 
gravitini remain massless, as expected (same is valid for dilatini).
On the other hand, gaugini masses are proportional to $f{}^{ab \hat c}$, 
so having  the same masses as their vector boson super-partners, vanishing 
at the 
self-dual enhancing points. Together with scalars  and 
vector bosons they fill up the ${\cal N}=4$ vector supermultiplet multiplet.

Let us argue that the discussion presented here for $d=4$ extends to other 
dimensions.
In fact, for half-maximal theories, the scalars form a coset $G/H = 
SO(d,d+n)/SO(d)\times SO(d+n)$ and are encoded in a coset representative 
$\mathcal{U}_{A}^{\bar A}=( {\cal U}_{A}{}^{a}; {\cal U}_{A}{}^{\hat 
I})$ as in 
\eqref{cosetrep} where $A$ is now a $G=SO(d,d+n)$-vector index, $a$ is a 
$SO(d)$-vector index and $\hat{I}$ is a $SO(d+n)$-vector index. The index 
$\hat{I}$ is  expressed in 
terms of spinor indices since fermions transform under $H=Spin(d)\times 
SO(d+n)$. From the full set of possible gaugings it is still possible to 
choose a subset parametrized by  an antisymmetric $G$-tensor, 
namely, $f_{ABC}$. For instance, 
in $d=4$ the full set of gaugings is parametrized by $\xi_{\alpha A}$ and 
$f_{\alpha ABC}$ ($\alpha=\pm$ is the electri-magnetic index)  and we have 
restricted to $\xi_{\alpha A}=0$ and 
$f_{+ABC}=f_{ABC}$, $f_{-ABC}=0$. The same applies in other 
dimensions\footnote{If we wanted it to also hold in $d=9$ and $d=8$ we must 
necessarily include vector multiplets, thus, $n\geq 1$. Otherwise, 
$f_{ABC}=0$, see \cite{Bergshoeff:2002nv,Dibitetto:2012rk}.}.
The fermion shift matrices would couple to scalars through the embedding tensor
and therefore they will necessarily have the same form as in 
\eqref{shiftmatrices} but where $i,j$ indices span the spinorial representation 
of $Spin(d)$.
 The reader should be aware that 
the actual mass terms of bilinear fermions are linear combinations of the above 
terms including scalars factors.

As before when the gaugings are the ones coming from an enhancement point of 
the 
string moduli space the structure constant (the fluxes) will take the values of 
the previous section. In these points, the gravitini shift matrices are zero 
and supersymmetry is preserved.  Away from the point of enhancement scalars 
vectors and fermions organized into a massive supermultiplet.

\section{Summary and outlook}
\label{sec:Summary and outlook}
In the present work  we have shown how  DFT can provide an 
interesting description of the gauge symmetry enhancing-breaking 
process  that occurs in the heterotic strings at specific points of moduli 
space. The construction relies on   previous ideas used 
to describe this process in the bosonic string case.
The three key ingredients encoding enhancing information are: a global  
$O(n_1,n_2)$ invariant gauged (super) gravity action, a scalar fluctuation 
expansion of a generalized scalar metric and the presence of generalized, 
moduli dependent, 3-form fluxes.

The heterotic effective action is obtained by choosing the global group  
$O(n,r)$ (where $n=\dim{G}_L$ is the dimension of the enhanced group at the 
fixed 
point and $r=10-d$ the number of compact dimensions)  and by identifying  
3-form fluxes $f_{ABC}(\Phi)$  with the internal momenta of the 
string.
Recall that indices are conveniently written as  $A=(a,\hat I)$ with  
$a=1,\dots 
n, \hat I=1,\dots r$. At a point of enhancement $\Phi_0$ the only non vanishing 
fluxes are those with only Left indices $f_{abc}(\Phi_0)$ reproducing the 
structure constants of the $G_L$ group. Away from this point, 
mixed indices give rise to non-vanishing fluxes $f_{ab\hat I}(\Phi)$. These 
mixed indices fluxes govern the vector boson masses, the scalar masses and the 
structure of the 
would be Goldstone bosons. It is worth emphasizing that this structure exactly 
matches the string theory results with the correct full dependence on moduli 
fields.

By invoking supersymmetry,  a fermionic sector can also be included. In 
particular we have shown that moving away from the enhancing point $\Phi_0$ 
produces the expected masses for gaugini partners of massive vector bosons 
while keeping supersymmetry unbroken.

Let us address some open questions.
In DFT the generalized fluxes appear as generalized Lie derivatives, involving 
internal coordinates $\mathbb Y^M$, of  generalized internal frame vectors 
$E_A(\mathbb Y)$. In a generic construction, the internal coordinates 
 transform in the vector representation of the global 
group $O(n_1,n_2)$, namely $M=1,\dots, n_1+n_2$ and the same is valid for the 
frame 
index $A$. However,  it appears that in order to reproduce the above fluxes,  
just a dependence on the ``true'' internal Left and Right $16+r+r=36-2d$ 
coordinates, associated to string  coordinates would be needed. In fact this 
was shown to be the case for some specific 
examples in \cite{agimnr,aamr} (see also \cite{Cagnacci:2017ulc}) for the 
bosonic string case. In a similar line of reasoning  a dependence on $\mathbb 
Y= (Y^I, 
y_L^{ I}, y_R^{\hat I})$ with $I=1,\dots 16; \hat I=1,\dots r$ would 
be expected. 
Therefore, the  tangent space here, spanned by $A$ would account for the 
gauge symmetry enhancement, associated to states with non vanishing KK 
momenta and windings, but the ``physical space''  
would be the string torus (including $\Gamma_{16}$).  The explicit  
construction for the heterotic string here  remains as an 
open question. 

Recall that our description is valid close to a given moduli point.When moving 
from one point of enhancement to a new point the dimension of the gauge group 
can 
drastically change and, therefore,  the dimension of the tangent space. Even
if, as stressed in \cite{aamr},  these tangent directions are not physical 
dimensions an explanation of how,  moving continuously 
from one point of enhancement to another  could lead to a discrete change in 
the 
number of these extra tangent dimensions is still lacking.  DFT description 
would presumably require the introduction of extra states, 
mimicking the string theory situation. Following the suggestions 
in \cite{aamr} this could be presumably achieved by considering a sort of 
generalized KK expansion on generalized momenta $\mathbb  L$ of the different 
fields coming into play. Thus, very schematically a vector boson 
corresponding to a charged generator would read\footnote{Similar expansions 
were considered in \cite{Aldazabal:2016yih} for the bosonic string case and 
for ${\mathbb L}^2=0$.}
\bea
A_{L\nu}(x, {\mathbb Y})
=\sum_{\mathbb L}
A_{L{\nu}}^{({\mathbb L})}(x) e^{i \mathbb{L}_{M} \mathbb{Y}^{ M}}\,\delta 
(\mathbb L^{ 2},1)
=\sum_{\mathbb L}
A_{L\nu}^{({\mathbb L})}(x) e^{i K.Y+ik_L.y_L+i k_R.y_R}\,\delta 
(\mathbb L^{ 2},1)\nn \, ,
\eea
where $K^I, k_L^m, k_R^m$ are functions of the moduli. Therefore,  when moving 
continuously along the moduli space, and for specific values of generalized 
momenta ${\mathbb L}$ in above sum, $k_R = 0 $ and the associated vector fields 
$A_{L\nu}^{({\mathbb L})}(x)$  become
massless. The neighborhood of each of such points is what our description 
would be capturing.

In order to address the  description of the enhancement process   we 
made a specific choice of generalized fluxes $f_{ABC}$ with  $A=(a,\hat I)$ by 
keeping just the indices leading to 
the enhanced gauge group structure constants at the enhancing moduli point and 
setting the other components to zero. However, it appears  interesting 
to explore the meaning of other possible components. In fact, we have already 
mentioned that, if we look at all indices running from $1,\dots r+n$, namely 
$a=i,\hat I=1,\dots r$ the corresponding fluxes encode the geometric and 
non-geometric (closed string) fluxes discussed in the literature\cite{fluxes}. 
In the six dimensional case, these fluxes span the ${\bf 220}$ representation 
of 
$O(6,6)$. Interestingly enough, the quadratic constraints \eqref{quadratic} 
mixing these fluxes with the gauge group ones would impose restrictions on 
the possible gauge groups. This is reminiscent of the Freed-Witten 
anomaly\cite{Freed:1999vc} cancellation requirements  discussed 
in \cite{Aldazabal:2008zza}, in the context of Type II string,   where such 
conditions where obtained  from quadratic constraints. Such mixings, in the 
heterotic string  Abelian case, were found also in \cite{Kaloper:1999yr}. 
Notice that there are still other fluxes components to be  considered that, in 
the context of Type II would correspond to mixings of open string indices with 
closed string ones. Heterotic/Type I duality could shed light on their possible 
interpretation.

\section*{Acknowledgments}
We thankD. Marqu\'es,  C. Nu\~nez and  A. Rosabal for useful 
discussions. 
This work was  supported
by  CONICET grant PIP-11220110100005  and   PICT-2012-513.
G. A. thanks the Instituto de F\'isica Te\'orica (IFT
UAM-CSIC) in Madrid for its support via the Centro de Excelencia Severo Ochoa
Program under Grant SEV-2012-0249.
\bigskip

\appendix

\section{Heterotic DFT}
\label{sec:Heterotic DFT}

In this section we briefly present some basic ingredients of DFT in the so 
called dynamical fluxes formulation. Details can be found in the 
references \cite{orDFT,reviewamn,edft,reviews,gm}.

In this formulation, the field degrees of freedom (metric, antisymmetric 
field, 1-forms) are encoded into generalized frame vectors  $E_{\bar {\cal 
A}}{}^{\cal M}$ 
that parametrize a coset  G/H where $G$ is a duality group.
Generalized metric is thus obtained from
\begin{equation}
 H=E_{\bar {\cal  A}}S^{\bar {\cal A} \bar {\cal B}}E_{\bar {\cal  B}}\, ,
 \label{genmetric}
\end{equation}
where  the $S_{\bar {\cal A}\bar {\cal B}}$ is given by
\be
 S_{\bar {\cal A} \bar {\cal B}} =
\left(\begin{matrix}s^{\bar a \bar b} & 0 \\ 0 & s_{\bar a \bar
b}\end{matrix}\right) \, ,\label{genFrame}
\ee
and $s_{\bar a \bar b}$ is the $d$-dimensional Minkowski metric. 

A scalar field $d$ incorporates the dilaton. Gauge invariance appears through a 
generalized Lie derivative ${\cal L}_{\xi}$. Transformations under this 
derivative lead to the dynamical fluxes.

The DFT action reads 
 \bea\label{ActionDFT}
  S \ = \ \int dX e^{-2d}\ {\mathbb{R}}\, ,
 \eea
 with
\bea\label{R}
	{\mathbb{R}} &=& {\cal F}_{\bar {\cal A}\bar {\cal B}\bar {\cal C}}\ 
{\cal 
F}_{\bar {\cal D}\bar
E\bar
F}\
	\left[\frac{1}{4} S^{\bar {\cal A}\bar {\cal D}} \eta^{\bar {\cal 
B}\bar 
E} \eta^{\bar
C\bar
F}-\frac{1}{12} S^{\bar {\cal A}\bar {\cal D}} S^{\bar {\cal B}\bar {\cal E}} 
S^{\bar {\cal C}\bar {\cal F}} - \frac
1 6
\eta^{\bar {\cal A}\bar {\cal D}}\eta^{\bar {\cal B}\bar {\cal E}}\eta^{\bar 
{\cal C}\bar {\cal F}}\right] \nn\\
&& + \ {\cal F}_{\bar {\cal A} } {\cal F}_{\bar {\cal B}}\  
\left[\vphantom{\frac 1
2}\eta^{\bar {\cal A} \bar {\cal B}} - S^{\bar {\cal A} \bar {\cal B}}\right]
\, ,
\eea 
where the dynamical fluxes ${\cal F}_{\bar {\cal A}\bar {\cal B}\bar {\cal C}}$ 
and ${\cal F}_{\bar {\cal A}}$ are defined in terms of the generalized Lie 
derivative $\cal L$ and 
vielbeins $E_{\bar {\cal A}}$ by
\begin{eqnarray}
{\cal F}_{\bar {\cal A}\bar {\cal B}\bar {\cal C}} &=& E_{\bar {\cal C} \cal M 
} 
{\cal L}_{E_{\bar {\cal A}}}
E_{\bar {\cal B}}{}^{\cal M} \label{DynFabc}\\
{\cal F}_{\bar {\cal A}} &=& - e^{2d}{\cal L}_{E_{\bar {\cal A}}} e^{-2d} \,.
\label{DynFa}
\end{eqnarray}

 The field  $d$ incorporates the dilaton field 
and transforms like a measure
\be
   {\cal L}_V e^{-2d}=\partial_P(V^{\cal P} e^{-2d})\, .
\ee

The indices ${\cal M}$ take values in the fundamental representation of the 
group $G$ 
whereas the flat indices $\bar {\cal A}$ run over $H$. Usually, motivated by 
the 
bosonic string construction the group $G$ is chosen to be $O(d,d)$ with $d$ 
the 
space-time dimensions. However the construction is more general and we can 
choose $G=O(d_L,d_R)$  generically containing 
$O(d,d)$ and $H=O(d_L)\times O( d_R)$. Indices are raised and 
lowered with a symmetric metric that can be chosen as 
 \be \label{etamd1d2}
\eta^{PQ}=\begin{pmatrix} 1_{d_L} & 0 \\ 0 & -1_{d_R} \end{pmatrix}.
\ee
A numerically similar matrix $\eta^{\bar {\cal A} \bar {\cal B}}$ is used for 
flat indices.

The dynamical fluxes depend on the generic  coordinates spanning a 
vector representation of $O(d_L,d_R)$. The above action is 
generically non invariant under generalized diffeomorphisms generated by the 
generalized Lie derivative unless  some constraints are  imposed. A 
consistent 
solution is given by a generalized Scherk Schwarz reduction where the frame 
 is split into a space-time dependent part and an internal one 
\cite{effective, reviewamn,gm}. The Lie derivative becoming a gauge 
transformation 
plus usual 
space-time diffeomorphisms.

Since we are interested in the description of the heterotic 
case we  perform a specif choice suitable for its description (see also 
\cite{reviewamn,hk,Hohm:2014sxa, Bedoya:2014pma}).  
Inspired by the coset structure presented in \eqref{dftembedding} we choose  
$d_L=d+n_c+26-d=d+n$ and $d_R=d+10-d=d+r$.

We also choose the fields to depend on the coordinates $\mathbb{X}^{\cal 
M}=\big(x_{\mu},\tilde{x}^{\mu},\mathbb{Y}^{M} \big)$ where 
 $\mathbb{Y}^{M}=(y_{L}^{m_L},y_{R}^{m_R} \big )$.  
 $M=(m_L,m_R)$ is an internal index with  $m_L=1,...,n$ and  
$m_R=1,...,10-d$ whereas  $\mu=1,...,d$.   Here $r=10-d$ is the 
number of  compact dimensions and $n=n_c+26-d$ is the number of extra 
directions needed to 
achieve the enhancement. Note that in section (\ref{sec:DFT rewriting}) the index $M$ is denoted as $A$ and $m_L= a$, $m_R=\hat{I}$. As usual, $\tilde{x}^{\mu}$ coordinates are just an 
artifact and can be dropped away or, in the DFT language, the strong 
constraint must be used on the space-time part.
 Explicitly the Scherk-Schwarz reduction ansatz reads
\beq \label{gss}
E_{\bar {\cal  A}}(x, y_{L},y_{R})= {\cal U}_{\bar {\cal  A}}{}^{\cal  {A}}(x) 
E'_{\cal  {A}}(y_{L},y_{R}) \ .
\eeq
The matrix ${\cal U}$ encodes the field content  in the effective 
theory, while
$E'$ is a generalized twist  that depends on the internal coordinates. All 
the 
dependence on the internal coordinates occurs through this twist.

 By introducing the splitting ansatz \eqref{gss} into the expression 
\eqref{genmetric} for the  generalized metric we can write
\begin{equation}
  H=S^{\bar {\cal A} \bar {\cal B}} {\cal U}_{\bar{\cal  A}}{}^{\cal 
A}(x) E'_{\cal A}(\mathbb 
Y) {\cal U}_{\bar{\cal B}}{}^{\cal B}(x) 
E'_{\cal B}(\mathbb Y)={\cal
H}^{\cal AB} (x)E'_{\cal A}(\mathbb Y) E'_{\cal B}(\mathbb Y) \, ,   
\end{equation}
where all the field dependence on space-time 
coordinates is encoded in
\begin{eqnarray}
{\cal H}^{\cal A B}(x) &=&
S^{\bar {\cal A} \bar {\cal B}} {\cal U}_{\bar{\cal  A}}{}^{\cal 
A}(x){\cal U}_{\bar{\cal B}}{}^{\cal B}(x) \, .
\label{scalarmatrix}
\end{eqnarray}
 In particular, when the indices  take internal  values 
$A,B=1,...,n+r$, the  matrix of ${\cal H}^{ A B}$ parametrizes the scalar 
content of the theory.

By restricting the expression for the generalized Lie derivative to the 
specific case of the twist it is found that 
\be
{\cal L}_{E^{\prime}_A}E^{\prime}_B=\frac12 \big[E^{\prime\ 
P}_A\partial_PE^{\prime\ M}_B-E^{\prime\ P}_B\partial_PE^{\prime\ 
M}_A+\eta^{MN}\eta_{PQ}\partial_NE^{\prime\ P}_AE^{\prime\ Q}_B \big]
\label{framederivative}
\ee
\beq
[E'_A,E'_B]={\cal L}_{E'_A}E'_B=f_{AB}{}^{K} E'_{K} \ ,
\label{structureconstants}
\eeq
 where here all indices are internal. The fluxes $f_{AB}{}^{K}$ of the 
generalised Scherk-Schwarz 
reduction must be 
constants and must  satisfy the constraints
\be \label{gsc}
f_{ABC}\equiv\eta_{AK} f_{AB}{}^L=f_{[ABC]}\, ,\qquad f_{[AB}{}^K
f_{K]C}{}^R=0\, 
\ee
in order for the algebra to close.

When replacing above results into the initial DFT action \eqref{ActionDFT}, the 
expression for the  gauged DFT action \eqref{actionDFTeff} is 
obtained.

Let us stress that a specific selection of values for the fluxes $f_{ABC}$, 
constructed out from the internal frame derivatives \eqref{framederivative} 
will be associated to a specific dependence on the generalized coordinates.
For instance, for the extreme case of a coordinate independent frame leads to 
an abelian 
compactification, and corresponds to a KK reduction, for instance. 

In particular, it was shown in 
\cite{agimnr,aamr,Cagnacci:2017ulc} that, at 
least for some 
cases, for the twists $E_A$ to reproduce the structure constants of the 
gauge group the twist only depends on the true internal coordinates of the 
torus. Thus,  for a circle compactification of the bosonic string 
the twist only depends on  the circle 
coordinate $y_L$ and its dual $y_R$. Explicitly\cite{aamr},
\bea\label{su2frame}\nn
E_{\pm}&=&  i\sqrt{\ap}( e^{\mp iw}  , \pm ie^{\mp iw} , 0;0)\qquad 
\qquad
E_{3}=  - i\sqrt{\ap}(0  ,0, 1  ; 0)\\
 \bar 
E_{\hat 3}&=& - i\sqrt{\ap}( 0, 0, 0;  1)\, ,
\eea
where $w= a_+y_L+a_-y_R, \bar w=a_-y_L+a_+y_R$\
and \begin{equation}
a_{\mp}= \sqrt{\frac{\ap}{2}}[\frac{1}{R}\mp\frac{1}{\tilde R}]\, .
\label{amp}
\end{equation}
It is easy to check that, by inserting this twist expression into 
\eqref{framederivative} and noticing that the only contributions to the 
derivatives come from $\partial_A=(0,\ 0,\ \partial_{ y_L}; \partial_{ 
y_R})$, the $SU(2)$ algebra is reproduced.
Here we just assume that there exists a choice of internal coordinates such 
that 
\eqref{structureconstants} leads to the desired gauge group structure 
constants and  leave the construction of the explicit twist for future 
work.

\section{Some Heterotic string basics}
\label{sec:Heterotic string basics}

We summarize here some string theory ingredients (that can be found in string 
books)  needed in the body of the 
article. We mainly concentrate in the $SO(32)$ string.

For a heterotic string compactified to $d$ space-time dimensions, 
Left and Right momenta are encoded in momentum 
\begin{equation}
        {\mathbb L}=(l_L, 
l_R)                                        
  \label{generalized}
  \end{equation}
defined on a self-dual lattice 
$\Gamma_{26-d,10-d}$ 
of signature $(26-d,10-d)$. By writing  $l_L^{\check{I}}=(K_L^I,k_{L,m})$ with 
$I=1,\dots,16$ and $m=1,\dots 10-d=r$, the moduli dependent momenta, read 
\bea
K_L^I&=& P^I+ RA_n^I \tilde p^n\\\nn
k_{L,m}&=&\sqrt{\frac{\ap}{2}}\big[\frac{p_m}{R}+(g_{mn}-B_{mn})\frac{\tilde 
p^n}{\tilde R}-P^I A_m^I-
\frac{R}{2}A_m^I A_n^I\tilde p^n\big]\\\nn
k_{R,m}&=&\sqrt{\frac{\ap}{2}}\big[\frac{p_m}{R}+(-g_{mn}-B_{mn})\frac{\tilde 
p^n}{\tilde R}-P^I A_m^I-
\frac{R}{2}A_m^I A_n^I\tilde p^n\big] \, ,
\label{leftrightmomenta}
\eea
where $g_{mn}, B_{mn}$ are internal metric and antisymmetric tensor 
components,  $A_m$ are Wilson lines and $ p_n$ and $\tilde p^n$ 
are integers corresponding to KK momenta and windings, respectively. $P^I$ are 
$Spin(32)$ 
weight components.

The  mass formulas for string states are 
  \begin{eqnarray}\nn
   \frac{\ap}2 m^{2}_{L}&=& \frac12 l_L^2+( N_B-1)=\frac12K_L^2+\frac12k_L^2+( 
N_B-1)\\
    \frac{\ap}2 m^{2}_{R}&=& \frac12k_R^2+ \bar N_B+\bar N_F+\tilde E_0\, ,
\label{LRstringmasses}
  \end{eqnarray}
 where $  N, \bar N$ are the number of string oscillators,  $\tilde 
E_0=-\frac12(0) $ for NS (R) sector  and the level matching condition is $ 
\frac12 m^{2}_{L}- \frac12 m^{2}_{R}=0$ or, in terms of above notation 
\begin{equation}
{\mathbb L}^2=\frac12 l_L^2- \frac12k_R^2=1-N_B-\bar N_B-\bar N_F-\tilde E_0 \, 
.
\end{equation} 
In particular, massless charged vectors correspond to ${\mathbb L}^2=1$. 
As is well known, there are $10-d+16$ Left gauge bosons corresponding to 
16 Cartan 
generators $\partial_z Y^I \tilde\psi^{\mu}$ of the original gauge algebra 
as well as $ 10-d$  KK Left gauge bosons 
coming from a Left combination of the metric and antisymmetric field 
$\partial_z Y^i \tilde\psi^{\mu}$. The $10-d$  Right combinations $\partial_z 
X^{\mu}\tilde\psi^{m}$ with 
$m=1,\dots 10-d$ generate the Right abelian group. These states have $k_R=0$ 
and $l_L=0$, with vanishing winding and KK momenta. 

Besides these states,  a number of 
different situations arises.
At generic points in moduli space  $k_R\ne 0$ and therefore there are no extra 
gauge bosons. The gauge group is then  $U(1)^{26-d}_L\times U(1)^{10-d}_R$. 
With  Wilson lines turned off,  for vanishing winding and 
momenta and  for $P^2=2$ the gauge  group   $SO(32)_L \times 
U(1)^{10-d}_L\times 
U(1)^{10-d}_R$ is obtained, as expected from 
KK reduction of the effective field theory.

For moduli points in $\Gamma_{16}$ such that 
\begin{equation}
 k_R=0, \qquad l_L^2=2 \, ,
 \label{masslesscharged}
\end{equation}
an enhancement of the Left gauge symmetry occurs, associated to the presence of 
massless gauge bosons $e^{i l_L. X_L(z)}\tilde\psi^{\mu}$. 
Actually, $l_L$  encodes the weight values associated to the charged generators 
of the enhanced 
algebra.
For specific values of moduli, momenta and windings a maximum enhancement to 
$SO(52-2d)_L\times U(1)^{10-d}_R$ 
gauge group can be obtained. Recall that the rank of the group is always 
$36-2d$ and the coset space is
\begin{equation}
\frac{O(26-d,10-d,\mathbb R)}{O(26-d,\mathbb R)\times 
O(10-d,\mathbb R)\times O(26-d,10-d,\mathbb Z)} ) 
 \label{oddstructure}
\end{equation}
of dimension $(26-d)(10-d)$. 



\begin{thebibliography}{98}

\bibitem{agimnr}
G.~Aldazabal, M.~Gra\~na, S.~Iguri, M.~Mayo, C.~Nu\~nez and J.~A.~Rosabal,
  ``Enhanced gauge symmetry and winding modes in Double Field Theory,''
  JHEP {\bf 1603}, 093 (2016)
  [arXiv:1510.07644 [hep-th]].
  
  
\bibitem{Cagnacci:2017ulc}
  Y.~Cagnacci, M.~Gra\~na, S.~Iguri and C.~Nu\~nez,
  ``The bosonic string on string-size tori from double field theory,''
JHEP {\bf 1706}, 005 (2017)  [ arXiv:1704.04242 [hep-th]].
  \bibitem{aamr}
G.~Aldazabal,  E. Andr\'es, M.~Mayo, and J.~A.~Rosabal,
 ``Gauge symmetry enhancing-breaking from a Double Field Theory perspective,''
  JHEP {\bf 1707} (2017)\\ 045 [arXiv:1704.04427 [hep-th]].

  
  
  
\bibitem{nsw}
  K.~S.~Narain, M.~H.~Sarmadi and E.~Witten,
  Nucl.\ Phys.\ B {\bf 279} (1987) 369.
  doi:10.1016/0550-3213(87)90001-0.
  
  K.~S.~Narain,
 ``New Heterotic String Theories in Uncompactified Dimensions $<$ 10,''
  Phys.\ Lett.\  {\bf 169B} (1986) 41.
  doi:10.1016/0370-2693(86)90682-9.

  
  
  \bibitem{Giveon:1994fu}
  A.~Giveon, M.~Porrati and E.~Rabinovici,
  ``Target space duality in string theory,''
  Phys.\ Rept.\  {\bf 244} (1994) 77
  [hep-th/9401139].
  
  \bibitem{ss}
  J.~Scherk and J.~H.~Schwarz,
  ``How to Get Masses from Extra Dimensions,''
  Nucl.\ Phys.\ B {\bf 153}, 61 (1979).
  
  \bibitem{effective}
    G.~Aldazabal, W.~Baron, D.~Marqu\'es and C.~Nu\~nez,
  ``The effective action of Double Field Theory,''
  JHEP {\bf 1111}, 052 (2011)
  [arXiv:1109.0290 [hep-th]].\\
  D.~Geissbuhler,
 ``Double Field Theory and N=4 Gauged Supergravity,''
  JHEP {\bf 1111}, 116 (2011)
  [arXiv:1109.4280 [hep-th]].
  
  
\bibitem{orDFT}
  C.~Hull and B.~Zwiebach,``Double Field Theory,''\\
  JHEP {\bf 0909}, 099 (2009) 
  [arXiv:0904.4664 [hep-th]].

  O.~Hohm, C.~Hull and B.~Zwiebach,
  ``Background independent action for double field theory,''
  JHEP {\bf 1007}, 016 (2010)
  [arXiv:1003.5027 [hep-th]].

  O.~Hohm, C.~Hull and B.~Zwiebach,
  ``Generalized metric formulation of double field theory,''
  JHEP {\bf 1008}, 008 (2010)
  [arXiv:1006.4823 [hep-th]].
  
  
  
  
  \bibitem{reviewamn}
G.~Aldazabal, D.~Marqu\'es and C.~Nu\~nez,
  ``Double Field Theory: A Pedagogical Review,''
  Class.\ Quant.\ Grav.\  {\bf 30} (2013) 163001
  [arXiv:1305.1907 [hep-th]].
  
\bibitem{gm} M.~Grana and D.~Marqu\'es,
  ``Gauged Double Field Theory,''
  JHEP {\bf 1204}, 020 (2012)
  [arXiv:1201.2924 [hep-th]].
  
  \bibitem{reviews}
  O.~Hohm, D.~Lust and B.~Zwiebach,
  ``The Spacetime of Double Field Theory: Review, Remarks, and Outlook,''
  Fortsch.\ Phys.\  {\bf 61}, 926 (2013)
  [arXiv:1309.2977 [hep-th]].
  
  
  
\bibitem{Schon:2006kz}
  J.~Schon and M.~Weidner,
  ``Gauged N=4 supergravities,''
  JHEP {\bf 0605}, 034 (2006)
  [arXiv:hep-th/0602024].
  
\bibitem{Samtleben:2008pe}
  H.~Samtleben,
  ``Lectures on Gauged Supergravity and Flux Compactifications,''
  Class.\ Quant.\ Grav.\  {\bf 25} (2008) 214002
  [arXiv:0808.4076 [hep-th]].
\bibitem{Trigiante:2016mnt}
  M.~Trigiante,
  ``Gauged Supergravities,''
   Phys.\ Rept.\  {\bf 680} (2017) 1
  [arXiv:1609.09745 [hep-th]].
  
\bibitem{Bergshoeff:1985ms}
  E.~Bergshoeff, I.~G.~Koh and E.~Sezgin,
  ``Coupling of Yang-Mills to N=4, D=4 Supergravity,''
  Phys.\ Lett.\  {\bf 155B} (1985) 71.
  doi:10.1016/0370-2693(85)91034-2    
  
  
\bibitem{fluxes}
J.~Shelton, W.~Taylor and B.~Wecht, ``Nongeometric flux
compactifications,''
  JHEP {\bf 0510} (2005) 085
  [hep-th/0508133].

G.~Aldazabal, P.~G.~C\'amara, A.~Font and L.~E.~Iba\~nez,
  ``More dual fluxes and moduli fixing,''
  JHEP {\bf 0605}, 070 (2006)
  [arXiv:hep-th/0602089].
  
  G.~Aldazabal, E.~Andr\'es, P.~G.~C\'amara and M.~Gra\~na,
  ``U-dual fluxes and Generalized Geometry,''
  JHEP {\bf 1011} (2010) 083
  [arXiv:1007.5509 [hep-th]].
 
  
   
\bibitem{Bergshoeff:2002nv}
E.~Bergshoeff, T.~de Wit, U.~Gran, R.~Linares and D.~Roest,
``(Non)Abelian gauged supergravities in nine-dimensions,''
JHEP {\bf 0210} (2002) 061
doi:10.1088/1126-6708/2002/10/061
[hep-th/0209205].

J.~J.~Fernandez-Melgarejo, T.~Ortin and E.~Torrente-Lujan,
``The general gaugings of maximal d=9 supergravity,''
JHEP {\bf 1110} (2011) 068
doi:10.1007/JHEP10(2011)068
[arXiv:1106.1760 [hep-th]].
  
  \bibitem{dftfermions}
  I.~Jeon, K.~Lee and J.~H.~Park,
  ``Incorporation of fermions into double field theory,''
  JHEP {\bf 1111} (2011) 025
  [arXiv:1109.2035 [hep-th]].
  
  I.~Jeon, K.~Lee and J.~H.~Park,
  ``Supersymmetric Double Field Theory: Stringy Reformulation of 
Supergravity,''
  Phys.\ Rev.\ D {\bf 85} (2012) 081501
   Erratum: [Phys.\ Rev.\ D {\bf 86} (2012) 089903]
  [arXiv:1112.0069 [hep-th]].
 
  

\bibitem{Jeon:2011cn} I.Jeon, K.Lee and J. H.Park,
``Stringy differential geometry, beyond 
Riemann'',  Phys.\ Rev.\ D {\bf 84} (2011) 044022
  doi:10.1103/PhysRevD.84.044022  [arXiv:1105.6294 [hep-th]].

  
\bibitem{Berman:2013cli}
  D.~S.~Berman and K.~Lee,
  ``Supersymmetry for Gauged Double Field Theory and Generalised 
Scherk-Schwarz 
Reductions,''
  Nucl.\ Phys.\ B {\bf 881} (2014) 369
  doi:10.1016/j.nuclphysb.2014.02.015
  [arXiv:1305.2747 [hep-th]].
  
   
  
 
  \bibitem{edft}
   D.~Geissbuhler, D.~Marqu\'es, C.~Nu\~nez and V.~Penas,
  ``Exploring Double Field Theory,''
  JHEP {\bf 1306}, 101 (2013)
  [arXiv:1304.1472 [hep-th]].
  
  
  
  

  
\bibitem{Dibitetto:2012rk}
G.~Dibitetto, J.~J.~Fernandez-Melgarejo, D.~Marques and D.~Roest,
``Duality orbits of non-geometric fluxes,''
Fortsch.\ Phys.\  {\bf 60} (2012) 1123
doi:10.1002/prop.201200078
[arXiv:1203.6562 [hep-th]].
  
 
  

  
 \bibitem{hk} O.~Hohm and S.~K.~Kwak,
  ``Double Field Theory Formulation of Heterotic Strings,''
  JHEP {\bf 1106}, 096 (2011)
  [arXiv:1103.2136 [hep-th]].
  
\bibitem{Hohm:2014sxa}
  O.~Hohm, A.~Sen and B.~Zwiebach,
  ``Heterotic Effective Action and Duality Symmetries Revisited,''
  JHEP {\bf 1502} (2015) 079
   [arXiv:1411.5696 [hep-th]].
  
\bibitem{Bedoya:2014pma}
    O.~A.~Bedoya, D.~Marqu\'es and C.~Nu\~nez,
  ``Heterotic $\alpha$'-corrections in Double Field Theory,''
  JHEP {\bf 1412}, 074 (2014)
  [arXiv:1407.0365 [hep-th]].

  
\bibitem{Freed:1999vc}
  D.~S.~Freed and E.~Witten,
  ``Anomalies in string theory with D-branes,''
   Asian J.\ Math.\  {\bf 3} (1999) 819
  [hep-th/9907189].

\bibitem{Aldazabal:2008zza}
  G.~Aldazabal, P.~G.~Camara and J.~A.~Rosabal,
  ``Flux algebra, Bianchi identities and Freed-Witten anomalies in F-theory 
compactifications,''
  Nucl.\ Phys.\ B {\bf 814} (2009) 21
   [arXiv:0811.2900 [hep-th]].
  
  \bibitem{Kaloper:1999yr}
  N.~Kaloper and R.~C.~Myers,
  ``The O(dd) story of massive supergravity,''
  JHEP {\bf 9905}, 010 (1999)
  [arXiv:hep-th/9901045].

  

  \bibitem{Aldazabal:2016yih}
  G.~Aldazabal, M.~Mayo and C.~Nu\~nez,
  ``Probing the String Winding Sector,''
  JHEP {\bf 1703} (2017) 096
  [arXiv:1611.04927 [hep-th]].




\end{thebibliography}
 \end{document}